\documentclass[a4paper,fleqn,usenatbib]{mnras}

\usepackage[T1]{fontenc}
\usepackage{ae,aecompl}

\usepackage{graphicx}	
\usepackage{amsmath}	
\usepackage{amssymb}	

\title[Giant planet migration via gap edge illumination]{Investigating the possibility of reversing giant planet migration via gap edge illumination}

\author[Hallam \& Paardekooper]{
P. D. Hallam\thanks{E-mail: p.d.hallam@qmul.ac.uk},
S.-J. Paardekooper
\\
Astronomy Unit, School of Physics and Astronomy, Queen Mary University of London, E1 4NS, UK
}

\date{Accepted XXX. Received YYY; in original form ZZZ}

\pubyear{2018}

\begin{document}
\label{firstpage}
\pagerange{\pageref{firstpage}--\pageref{lastpage}}
\maketitle

\begin{abstract}
A massive planet in a protoplanetary disc will open a gap in the disc material which acts as a transition between Type I and Type II planetary migration. Type II migration is slower than Type I migration, however it is still desirable to slow down Type II migration to allow gas giant planets with semi-major axis in the range $5-10$AU to exist, similarly to our Solar system. We investigate a method of slowing down and reversing Type II migration by heating the outer gap edge due to incident radiation from the central star. Using an approximate vertically averaged heating method we find that Type II migration can be slowed or in extreme cases reversed if we assume near maximum allowed irradiation from the central star. Therefore, we believe this is a very interesting phenomenon that should be investigated in greater detail using three dimensional hydrodynamic and radiative transfer simulations.
\end{abstract}




\section{Introduction}
Planets form in protoplanetary discs by accretion of gas and dust particles and will excite density waves in the disc material that transport angular momentum away from the planet. This angular momentum is deposited in the disc, exerting a torque on the disc material. If the angular momentum is deposited close to the planet \citep{Lin&Papaloizou1993,Bryden1999} and the resultant torque is stronger than the viscous diffusion torque of the disc material \citep{Lin&Papaloizou1979,Goldreich&Tremaine1980,Takeuchi1996,Crida2006} then a gap is formed around the location of the planet. These two conditions are known as the thermal and viscous criteria for gap opening. It has also been shown that small mass planets can open gaps at a greater distance from the planet, where the excited density wave shocks \citep{Goodman&Rafikov2001, Rafikov2002}.

The interaction between the planet and the disc plays an important role in the evolution of the planet's orbital radius. There are two main regimes of planetary migration, Type I and Type II, which are differentiated between by the mass of the planet and as a result, the presence of a gap in the disc material. Type I migration occurs for planets that are not massive enough to open a gap in the disc, and migrate due to the Lindblad and corotation torques acting upon them \citep{Goldreich&Tremaine1980}. The corotation torque depends on the magnitude of the thermal and viscous diffusion in the disc \citep{Masset2001,Paardekooper2010,Paardekooper2011} and these can alter the speed of migration. Type II migration occurs when a planet is massive enough to open a gap in the disc. The low density area surrounding the planet slows down the Type I migration, and the planet now migrates with the gap on the viscous evolution timescale of the disc \citep{LinPapaloizou1986}. Hence Type I migration is considerably faster than Type II migration \citep{Ward1997}. Despite this reduction in migration rate, it has been found that the predicted timescale for Type II migration is shorter than the lifetime of the disc, suggesting that most planets undergoing Type II migration will migrate into and be absorbed by the central star \citep{Hasegawa&Ida2013}. Observational evidence dictates that gas giant planets are more common at orbital distances $R > 1 \textrm{AU}$ \citep{Mayor2011, Cassan2012,Fressin2013,Santerne2016}, which would not be the case if this predicted timescale was correct. Current models estimate that for giant planets to survive migration they must form at distances $R>20\textrm{AU}$ \citep{Coleman&Nelson2014}. Hence it is implied that something must be limiting the migration speed of a planet in the Type II migration regime \citep{Nelson2000}.
 
\begin{figure}
	\includegraphics[width=\columnwidth]{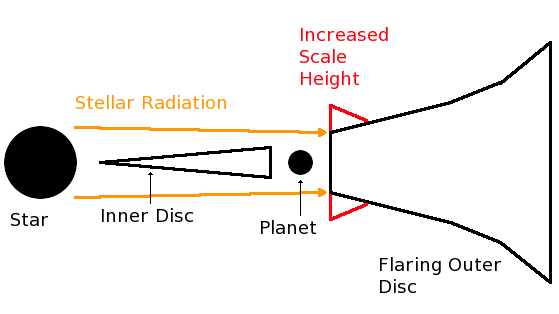}
	\vspace*{-5mm}
    \caption{A diagram showing the process by which the outer gap edge is heated by the star. This is the method in which the torque contribution from the outer disc may be reduced and hence the net torque on the planet may become more positive.}
    \label{Fig:Diagram}
\end{figure}

The argument for classical Type II migration relies on the absence of material surrounding the planet, a result of the torque exerted on the disc by the massive planet clearing the area around it and forming a gap. This is clearly the case in one dimensional simulations, such as those in \cite{LinPapaloizou1986}, however extending this to higher dimensions this is found not to be the case \citep{Kanagawa2015,Hallam&Paardekooper2017}. It can easily be seen from two or three dimensional simulations that disc material can be present within the gap, and can cross the gap on horseshoe orbits. Hence we can only view classical Type II migration as an idealised case. The result is a deviation from the classical Type II migration rate, the viscous evolution timescale of the disc. The migration rate can now be significantly faster or slower than the viscous evolution timescale of the disc and is dependent on the planet and disc parameters \citep{Duffel2014}. \cite{Durmann&Kley2015} found that only for small disc masses $M_d/M_J < 0.2$ is the migration rate slower than the classical rate. Therefore, we investigate a new method of slowing the rate of planet migration in the Type II regime. We propose that radiation from the central star is incident on the outer edge of gap formed by the giant planet, visible as a result of the flaring of the disc. This incident radiation heats the outer disc, increasing the scale height in this region. This process is outlined in Figure \ref{Fig:Diagram}. As the one-sided torques scale with the aspect ratio as $h^{-3}$, the result of this may be a lowered contribution to the net torque on the planet from the planet's outer wake, which in turn causes the net torque on the planet to become more positive. As the rate of migration is proportional to the magnitude of the torque on the migrating planet and the direction given by the sign of the torque, this process could slow the planet's migration. Observational signatures of irradiated gap edges have been studied by \cite{Jang-Condell2013}.

This paper is arranged as follows. In Section \ref{Sec:Basic_Eq} we derive the relevant equations solved to simulate disc evolution. In Section \ref{Sec:NumSet} we discuss the code used and the numerical setup for our simulations. In Section \ref{Sec:Just} we justify the choice of parameters used in our simulations. In Section \ref{Sec:Inst} we discuss the onset of instabilities in our simulations and how we go about avoiding them. In Section \ref{Sec:Res} we present our results of heating the gap edge and the resultant dependence of the net torque on previously justified parameters. In Section \ref{Sec:Disc} we discuss our results in the context of prior work while justifying assumptions made and postulating any impact they may have. Finally, in Section \ref{Sec:Conc} we present our conclusions.

\section{Basic Equations}\label{Sec:Basic_Eq}
In this paper we perform a parameter study in two dimensions to outline a region for which there is the possibility of reversing the direction of Type II migration. While the heating of the outer disc is a radiative hydrodynamic process which is clearly a three dimensional phenomenon, we choose to represent this phenomenon in a vertically averaged approach outlined in Section \ref{SubSec:Heating}. A full radiation-hydrodynamic parameter study in three dimensions is beyond current computational resources and an approximate study such as presented here can serve as a guide for future more realistic simulations.

\subsection{Two dimensional protoplanetary disc}\label{Sec:2Ddisc}

The continuity equation for the evolution of a protoplanetary disc's surface density, $\Sigma$, is 

\begin{equation}\label{eq:Cont}
	\frac{\partial\Sigma}{\partial t} + \nabla\cdot (\Sigma \mathbf{v}) = 0,
\end{equation}
where $\mathbf{v}$ is the velocity field. We simulate the evolution of a protoplanetary disc's surface density due to the presence of a planet by solving the two dimensional Navier-Stokes equation for the motion of the disc planet system,

\begin{equation} \label{eq:Nav_Stokes}
	\Sigma\left(\frac{\partial\mathbf{v}}{\partial t} + \mathbf{v}\cdot\nabla\mathbf{v}\right) = -\nabla P -\nabla\cdot\mathbf{T} - \Sigma\nabla\Phi,
\end{equation}
where $\mathbf{T}$ is the Newtonian viscous stress tensor, $P$ is the pressure and $\Phi$ is the gravitational potential of the planet and star system. We solve the energy equation,

\begin{equation} \label{eq:Energy}
	\frac{\partial e}{\partial t} + \nabla\cdot e\mathbf{v} = -P\nabla\cdot(\mathbf{v}) + \mathcal{C},
\end{equation}
with an equation of state $P =(\gamma - 1)e$, where $e$ is the volumetric internal energy and $\gamma$ is the ratio of specific heats. $\gamma = c_P/c_V = 1.4$ for a diatomic gas. $\mathcal{C}$ is a cooling term, in which we introduce a thermal relaxation timescale, $\tau$, as discussed in Section \ref{SubSec:Heating}. A cylindrical coordinate system is used, such that $\mathbf{v} = (v_R,R\Omega)$ where $v_R$ and $\Omega$ are the radial and angular velocities at a given radius. The planet is held on a fixed circular orbit.

\subsection{Heating the outer gap edge}\label{SubSec:Heating}
\begin{figure}
	\includegraphics[width=\columnwidth]{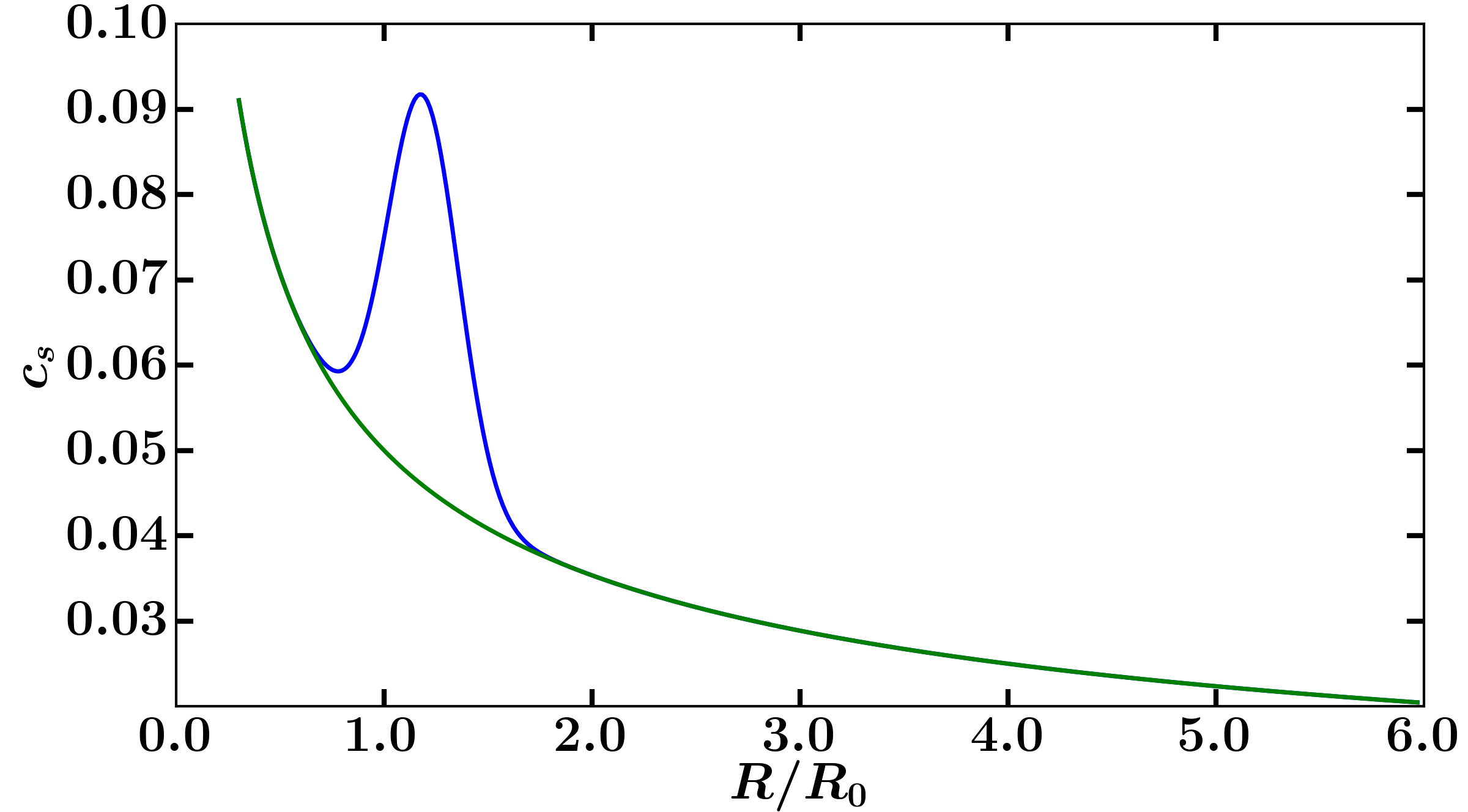}
	\vspace*{-5mm}
    \caption{Azimuthally averaged initial sound speed profiles for a disc containing a Gaussian with $A = 2.0$ and $\sigma = 0.17$, described by Equation \ref{eq:Gaussian_Profile} and a disc containing no Gaussian in the sound speed profile. Discs setups are described in more detail in Section \ref{Sec:NumSet} and the parameters of this Gaussian are discussed in more detail in Section \ref{SubSec:Stand}.}
    \label{Fig:SoundSpeeds}
\end{figure}

In order to attempt to reduce the rate of Type II migration we consider the impact on the net torque on the planet if the outer gap edge was directly heated by radiation from the central star. This may reduce the torque contribution from the outer wake of the planet, while leaving the contribution from the inner wake largely unchanged. To represent this in our simulations we impose a Gaussian distribution in the sound speed profile of the disc, $c_s$, of the form

\begin{equation} \label{eq:Gaussian_Profile}
	c_s = \left(1.0+(A-1.0)e^{\left(-\frac{\left((R-R_G\right)^2}{2\sigma^2}\right)}\right)h\left(\frac{R}{R_0}\right)^f\sqrt{\frac{GM_*}{R}},
\end{equation}
where $A$ is the factor by which the sound speed increases at $R = R_G$, $R$ is the radial position within the disc, $R_G$ is the location of the Gaussian peak, $\sigma$ is the standard deviation of the Gaussian, $R_0$ is the location of the planet, $f$ is the flaring index and $M_*$ is the mass of the star. This represents the stellar heating of the outer gap edge, resulting in an increase in sound speed at that location. This changes both the sound speed and aspect ratio of the disc at this location. The impact a Gaussian has on the sound speed profile of the disc is shown in Figure \ref{Fig:SoundSpeeds}. In this paper we will be investigating the effect of the shape of this Gaussian on the net torque on the planet.

We deviate from a purely adiabatic simulation by applying a thermal relaxation function at constant density, given by

\begin{equation} \label{eq:Cooling}
	\mathcal{C} = -\frac{e-e_i\frac{\Sigma}{\Sigma_i}}{\tau}
\end{equation}
where $e$ is the volumetric internal energy, $\tau = 2\pi/\Omega_0$ is the thermal relaxation timescale, $\Omega_0$ is the angular velocity at the location of the planet and the subscript $i$ denotes initial and therefore equilibrium values. As $\tau$ is a constant this means the rate of cooling in the inner disc will be slower than cooling in the outer disc when compared with the orbital timescale at that location. This does not affect our results, as we are interested in cooling at and around the location of the planet. This is then applied using implicit time stepping to cool the disc in the form 

\begin{equation} \label{eq:Cooling2}
	e^{n+1} = e_i\frac{\Sigma^n}{\Sigma_i} + \frac{e^n-e_i\frac{\Sigma^n}{\Sigma_i}}{1+\frac{\Delta t}{\tau}}
\end{equation}
where the superscript $n$ denotes the timestep at which the relevant field is evaluated. 

\subsection{Total Torque}\label{SubSec:TorqueCalc}

The total torque per unit mass on the planet was calculated using

\begin{equation}
	\mathcal{T}_m = \frac{1}{M_p}\int_{R_{\textrm{Min}}}^{R_{\textrm{Max}}}\int_{-\pi}^{\pi}\Sigma(R,\phi)\frac{d\Phi_p}{d\phi}d\phi RdR
\end{equation}
where $R_{\textrm{Min}}$ and $R_{\textrm{Max}}$ are the minima and maxima of the radial domain respectively, and $\Phi_p$ is the gravitational potential of the planet, given by

\begin{equation}
	\Phi_p = -\frac{GM_p}{\sqrt{R^2 + R_0^2 - 2RR_0\cos(\phi-\phi_0) + s^2}},
\end{equation} 
in which $s = 0.6H_0$ is the smoothing length, $R_0$ is the location of the planet, $H_0$ is the scale height at the location of the planet and $\phi_0$ is the azimuthal location of the planet. The contribution of the planet envelope within $0.8$ Hill Radii is not included in the calculation of torque on the planet, similarly to \cite{ Crida2008,Durmann&Kley2015}. Our cut off is applied before force calculation, and ramps sinusoidally back from $0$ to the default value at that radii,

\begin{equation}
	\mathcal{H}_\textrm{c} = \begin{cases}
			0 & \textrm{if} \>\>r/r_{\textrm{Hill}} < 0.8 \\
			1 & \textrm{if} \>\>r/r_{\textrm{Hill}} > 1.0 \\
			\sin^2\left({\frac{\pi}{0.4}\left(\frac{r}{r_{\textrm{Hill}}} - 0.8\right)}\right) &\textrm{otherwise},
	      \end{cases}
\end{equation}
where $r$ is the distance from the planet. Then,

\begin{equation}
	F_{\textrm{cut-off}} = F\mathcal{H}_\textrm{c}
\end{equation}

We expect a change in sound speed to impact the net torque on the planet. For a low mass planet, the one sided torques in the disc are given by \citep{Ward1997}

\begin{equation} \label{eq:OneSideTorque}
	\mathcal{T_\textrm{os}} = C\left(\frac{M_p}{M_*}\right)^2\left(\frac{1}{h}\right)^3\Sigma_0 R_0^4\Omega_0^2,
\end{equation}
where $C$ is a constant that is positive for the inner disc and negative for the outer disc and $\Sigma_0$ is the unperturbed surface density at $R = R_0$. In an unperturbed disc the one sided torques will mostly cancel, resulting in a small net torque on the planet. \cite{Tanaka2002} has shown that for a two dimensional disc containing a low mass planet the overall net torque on the planet is given by

\begin{equation}\label{eq:NetTorque}
	\mathcal{T} = -(1.160+2.828\alpha_\textrm{sl})\left(\frac{M_p}{M_*}\frac{1}{h}\right)^2\Sigma_0 R_0^4\Omega_0^2,
\end{equation}
where $\alpha_\textrm{sl}$ is the negative of the radial surface density gradient. As $h \ll 1$, therefore $\mathcal{T} \ll \mathcal{T_\textrm{os}}$. If the temperature and therefore sound speed in the outer disc is greater than that of the inner disc Equation \ref{eq:OneSideTorque} implies that the inner torque will be favoured. Despite Equations \ref{eq:OneSideTorque} and \ref{eq:NetTorque} pertaining to small planets, this is expected to hold for higher mass planets.

\section{Numerical Setup} \label{Sec:NumSet}
Our simulations are run in two dimensions using the FARGO3D code. This is a magnetohydrodynamic code designed to simulate disc evolution in one to three dimensions by solving the hydrodynamic equations of motion. FARGO3D is a good choice for investigating gap opening planets as it uses a C to CUDA translator to allow simulations to be run on Graphics Processing Units (GPUs). GPUs have limited memory, but decrease computational time significantly, allowing moderate resolution simulations to be run for extended periods of time. For more details see \cite{FARGO3D}.

Our disc and planet parameters remain constant between simulations, the variables modified pertain to the applied Gaussian in the sound speed profile only. Hence the setup described below is the standard disc-planet setup used in all simulations. We use a $q = M_p/M_* = 0.001$ planet, which is equivalent to one Jupiter mass around a Solar type star. This is large enough to undergo gap formation and enter the Type II migration regime. Our disc extends over a radial domain of $0.3 \leq R/R_0 \leq 6.0$ and azimuthal domain $-\pi \leq \phi \leq \pi$ with $285$ by $596$ cells respectively, using logarithmic cell spacing in the radial direction. This radial range was chosen to ensure no torque contribution from the outer disc was suppressed when calculating the net torque on the planet. The initial surface density gradient is $\Sigma_{\textrm{int}} \propto R^{-1/2}$ with $\Sigma_0 = 2.67\times 10^{-3}$ at $R_0 = 1.0$ and we use a constant viscosity $\nu = 7.5\times 10^{-6}$ corresponding to a dimensionless viscosity $\alpha = 3\times 10^{-3}$ using the alpha prescription $\nu = \alpha c_s H$ of \cite{ShakuraSunyaev1973}. We use a constant aspect ratio $h = H/R = 0.05$. Reflecting boundary conditions were used with wave killing zones in the regions $0.3 \leq R/R_0 \leq 0.585$ and $5.43 \leq R/R_0 \leq 6.0$ similarly to \cite{DeValBorro2006}, meaning that excited waves are damped before reaching the edge of the simulation. The planet is held on a fixed circular orbit with no migration or disc self gravity accounted for. The migration of the planet is predicted from the net torque on the planet, which is determined by allowing the simulation to run for $600$ orbits, and averaging the net torque on the planet over the last $100$ orbits. Gap edge heating from the star is represented by a Gaussian in the sound speed profile of the disc, centred at the outer edge of the planet gap. Deviation from a purely adiabatic simulation is a result of the applied thermal relaxation function, Equation \ref{eq:Cooling}.

\section{Justification of Gaussian Parameters}\label{Sec:Just}
While the parameters pertaining to the disc and the planet have been discussed in Section \ref{Sec:NumSet}, in this section we will justify the selection of the Gaussian parameters used in our simulations. Here we must reiterate what the presence of the Gaussian in the sound speed profile is intending to represent, the heating of the gap edge due to the incident radiation from the star. The important parameters are all contained within Equation \ref{eq:Gaussian_Profile}, namely the amplitude of the a Gaussian, $A$, the location of the Gaussian peak $R_G$ and the full width at half maximum of the Gaussian, $\mathcal{W}$, from which $\sigma$ is determined.

The easiest to identify of these parameters is $R_G$, as this will obviously occur at the gap edge and can be roughly determined from running a very basic simulation of a disc plus a planet (no Gaussian applied yet). The specific location of the gap edge, however, is more difficult to determine, as its exact definition would be dependent on the point at which the disc material becomes opaque to the stellar radiation. As exact calculations involving disc opacity are beyond the scope of this paper, we instead consider the outer slope of the gap in the surface density profile, from which we decide that $R_G = 1.20 - 1.40$ are a reasonable range within which the disc is heated. We also find that the change in $c_s$ due to the Gaussian does not alter the location of the gap edge, hence these values of $R_G$ are always located on the gap edge.

As the Gaussian acts as a local increase in the sound speed profile by a factor $A$, in order to predict a rough value of $A$ we must estimate the sound speed increase due to the heating of the gap edge. This means we must first determine how much the temperature of the gap edge increases due to stellar heating, and then how this corresponds to an increase in sound speed. Therefore, we must calculate the unperturbed equilibrium temperature of the disc, and then the increased temperature due to the gap. Using the results of \cite{Chiang&Goldreich1997} for geometrically flat discs, we use the equation

\begin{equation}
	T_\textrm{e} \approx	\left(\frac{2}{3\pi}\right)^{\frac{1}{4}}\left(\frac{R_*}{R}\right)^\frac{3}{4}T_*
\end{equation}
where $T_\textrm{e}$ is the effective temperature of the disc and $R_*$ and $T_*$ are the stellar radius and temperature respectively.

To calculate the temperature of the heated gap edge we make the further assumption that there is no planet and no inner disc. This allows us to treat the gap edge as the inner edge of the disc from which we can apply the results of \cite{Dullemond2001}. While clearly this is not the exact case we are investigating, the presence of an inner disc could be represented by an efficiency factor, blocking some of the radiation incident on the gap edge in our simplified case. This means that the case we present here represents the maximum possible gap edge irradiation for a given set of stellar parameters. Therefore to calculate the temperature of the heated gap edge we can use

\begin{equation}\label{Eq:T_G}
	T_G \approx	\left(\frac{L_*}{4\pi R_G^2\sigma_{sb}}\right)^{\frac{1}{4}}
\end{equation}
where $T_G$ is the effective temperature of the gap edge, $L_*$ is the luminosity of the star and $\sigma_{\textrm{sb}}$ is the Stefan-Boltzmann constant. In determining Equation \ref{Eq:T_G} we also make the very reasonable assumption that $R_G \gg H_G$, namely the aspect ratio at the edge of the gap is much smaller than the radial location of the gap edge.

Now from these we have a ratio of temperatures, $T_G/T_\textrm{e}$ which we need to convert into a ratio of sound speeds. The temperature and sound speed are related by the pressure

\begin{equation}
	P = \frac{\Sigma R_\textrm{m} T}{\mu} = c_s^2\Sigma
\end{equation}
where $R_\textrm{m}$ is the molar gas constant and $\mu$ is the mean molecular weight. Therefore,

\begin{equation}\label{Eq:Ratio}
	A = \frac{c_G}{c_\textrm{e}} = \left(\frac{T_G}{T_\textrm{e}}\right)^\frac{1}{2},
\end{equation}
so the ratio of sound speeds, and therefore the parameter $A$ depends on the the root of the ratio of the temperatures. Using Equation \ref{Eq:Ratio} we can see that the $A$ has a weak dependence of $R_G$, such that $A \propto R_G^{1/8}$, however this is strongly dependent on the assumed disc geometry. Hence we estimate the value of $A$ at $R_G = 1.0$AU for a number of different stellar classifications. We include a T-Tauri star in this range of stellar classifications, using the stellar parameters found in \cite{Dullemond2001} for our T-Tauri calculation.

This is presented in Table \ref{Table:Stars}, but the general conclusion is that $A$ ranges from around $1.8 - 2.5$ and increases as the star becomes smaller, cooler and less luminous. When we choose a value of $A$ for our Gaussian, we can consider it the maximum irradiation from the star, in which case the values presented in Table \ref{Table:Stars} would be the $A$ we select, or we could consider it as a reduced magnitude of irradiation from a star that would provide a greater value of $A$ at maximum irradiation. For example for a T Tauri star the maximum gap edge irradiation from Table \ref{Table:Stars} is $2.18$, while an $A = 2.00$ or $1.75$ could still represent a T Tauri star at near maximum gap edge irradiation, or a different classification of star at maximum irradiation.

\begin{table}
\caption{Values of the parameter $A$ for different stellar classifications}
\centering
\begin{tabular}{c|c|c|c|c}
\hline
Classification & $T_*/K$ & $R_* /R_{\textrm{sun}}$ & $L_*/W$ & $A$ \\
\hline
O & 30000 & 6.60 & $1.2\times10^{31}$ & 1.87 \\
B & 20000 & 4.20 & $9.8\times10^{29}$ & 1.99 \\
A & 8750 & 1.60 & $5.2\times10^{27}$ & 2.24 \\
F & 6750 & 1.30 & $1.2\times10^{27}$ & 2.30 \\
G (Sun) & 5780 & 1.00 & $3.8\times10^{26}$ & 2.37 \\
K & 4500 & 0.83 & $9.8\times10^{25}$ & 2.43 \\
M & 3700 & 0.70 & $3.2\times10^{25}$ & 2.49 \\
T-Tauri  & 3800 & 2.54 & $2.9\times10^{26}$ & 2.18\\

\hline
\end{tabular}
\label{Table:Stars}
\end{table}

Recalling that this is evaluated at $R_G = 1.0$AU rather than the actual radial location of the gap edge, the exact location of heating depends greatly on the opacity of the disc material and the inner disc could act as an efficiency factor to block some of the incident radiation, we can only call this a rough estimate. Despite this it gives us a good range in the parameter $A$ for us to explore.

The next parameter, $\mathcal{W}$, is more difficult to select. Theoretically $\mathcal{W}$ represents the redistribution of temperature from the primary heating location ($R_G$), however this will be dependent once again on the opacity of the material and also the density of material, which will be modified by the presence of a gap opening planet. We select $\mathcal{W}$ based on the constraint discussed in Section \ref{Sec:Inst}, that a too narrow Gaussian leads to the excitement of instabilities. This sets a lower limit in our selection of $\mathcal{W}$. We explore a range of Gaussians with $\mathcal{W}$ above this, however we know that if we make the Gaussian too wide the heating will spread into the corotation region and the inner wake of the planet. If the torque contribution from the outer wake of the planet is also reduced by the Gaussian then the overall impact on the net torque due to the gap edge heating will be reduced. Hence we believe this will act to reduce the parameter space for which we can successfully slow Type II migration. The region covered by the Gaussian corresponds to a "puffed up" gap edge in the disc and hence regions behind this will be self-shadowed from stellar radiation. As a result the temperature in these regions will be reduced. This effect is an extra complication, in addition to not knowing the exact heating profile.

\section{Instability Study}\label{Sec:Inst}

Implementing a Gaussian distribution in the sound speed profile of the disc often results in the excitation of Rayleigh or Rossby Wave instabilities, and subsequent destabilisation of the Gaussian, resulting in a very different shape from the input equilibrium profile. We study the parameter space for which an input Gaussian remains stable and use this to ensure stable Gaussians are implemented when investigating the net torque on the planet. We do this by investigating the behaviour of a Gaussian in the sound speed profile of a disc with setup described by Sections \ref{Sec:Basic_Eq} and \ref{Sec:NumSet}, however these simulations contain no planet. We investigate these setups by both simulating the time evolution of the disc and by numerically analysing instability criteria. There are two potential instabilities excited by the implementation of a Gaussian in the sound speed profile, these are the Rayleigh Instability and the Rossby Wave Instability, which are addressed in Section \ref{SubSec:RI} and Section \ref{SubSec:RWI} respectively.

\subsection{Rayleigh Instability}\label{SubSec:RI}

For a disc to be Rayleigh stable the Rayleigh stability condition,

\begin{equation}\label{Eq:Rayleigh}
		\frac{d}{dR}\left(R^2\Omega\right) > 0,
\end{equation}
must be satisfied \citep{Chandrasekhar1961}. This is satisfied for a disc containing no planet undergoing Keplerian rotation, however modifying the sound speed profile opens up the possibility for this criterion to be violated. The angular velocity is given by

\begin{equation}
		\Omega^2 = \Omega_K^2 + \frac{1}{R\Sigma}\frac{\partial P(c_s)}{\partial R},
\end{equation}
so clearly a change in the sound speed profile will impact the stability of the system. For large enough deviations from Keplerian angular velocity, $\Omega_K$, we would expect to find the disc Rayleigh unstable. We test this stability criterion for the sound speed profile given in Equation \ref{eq:Gaussian_Profile}, with $R_G = 1.20$, and find a range of Gaussians for which the Rayleigh criterion is satisfied. It is found that for a given $A$, a $\mathcal{W}$ exists beyond which Gaussians will be Rayleigh stable. This information is presented in the second column of Table \ref{Table:Rossby}. 

Running simulations of discs with Gaussians of these parameters we find that the Gaussians still go unstable even when the Rayleigh criterion is satisfied, however as expected an increase in $\mathcal{W}$ keeps the Gaussian stable for longer. This leads us to conclude that the Rayleigh instability is not the only instability we must be concerned with. 

\subsection{Rossby Wave Instability}\label{SubSec:RWI}

In addition to the Rayleigh criterion from Equation \ref{Eq:Rayleigh} a Gaussian must also satisfy the Rossby Wave criterion to be stable to the Rossby Wave instability. Similarly to the Rayleigh criterion there is a critical value of $\mathcal{W}$ for a given $A$ beyond which the Gaussian will be Rossby stable. This is in all cases a larger $\mathcal{W}$ than that which satisfies the Rayleigh criterion, meaning that Gaussians with very small $\mathcal{W}$ below those listed in the second column of Table \ref{Table:Rossby} will be both Rossby and Rayleigh unstable.

A local maxima or minima in the radial profile of the key function,

\begin{equation}\label{Eq:KeyFunc}
	\mathcal{L}(R) = \mathcal{F}S^\frac{2}{\gamma}
\end{equation}
is required for the Rossby Wave instability to occur \citep{Lovelace1999}. Here $\mathcal{F}^{-1} = \mathbf{\hat{z}}\cdot(\nabla\times\mathbf{v})/\Sigma$ is the potential vorticity and $S = P/\Sigma^{\gamma}$ is the entropy. As both potential vorticity and entropy have dependence on the sound speed, again we can see that modifying the sound speed profile will impact the Rossby stability of the system. In addition to the local maxima/minima, an instability only occurs if a threshold in the variation of the key function is reached and exceeded. This threshold is non-trivial to predict and is dependent on the input parameters of the simulation \citep{Lovelace1999, Ono2016}. We can study the Rossby stability of our Gaussians simply by simulating their time evolution in a disc and observing the results in the two dimensional sound speed distribution, the azimuthally averaged sound speed distribution and the radial velocity distribution as the presence of the Rossby Wave instability is clearly visible from these. Hence we can easily find a $\mathcal{W}$ for a given $A$ at which the Gaussian becomes Rossby stable, as at this point there will be no non-axisymmetric structure in these distributions, and the sound speed profile will be unchanged from the initial condition to the final output. This information is presented in the third column of Table \ref{Table:Rossby}.  

\begin{table}
\caption{$\mathcal{W}$ limits for Rossby and Rayleigh stable Gaussians of height $A$}
\centering
\begin{tabular}{c|c|c}
\hline
	& Rayleigh Stable & Rossby and Rayleigh Stable \\
$A$ & $\mathcal{W}$ & $\mathcal{W}$ \\
\hline
1.50 & 0.20 & 0.25 \\
2.00 & 0.30 & 0.40 \\
2.15 & 0.35 & 0.45 \\
2.25 & 0.35 & 0.50 \\
2.50 & 0.40 & 0.60 \\
\hline
\end{tabular}
\label{Table:Rossby}
\end{table}

\subsection{Instabilities excited by the presence of the planet}\label{SubSec:PlanetInst}
\begin{figure}
	\includegraphics[width=\columnwidth]{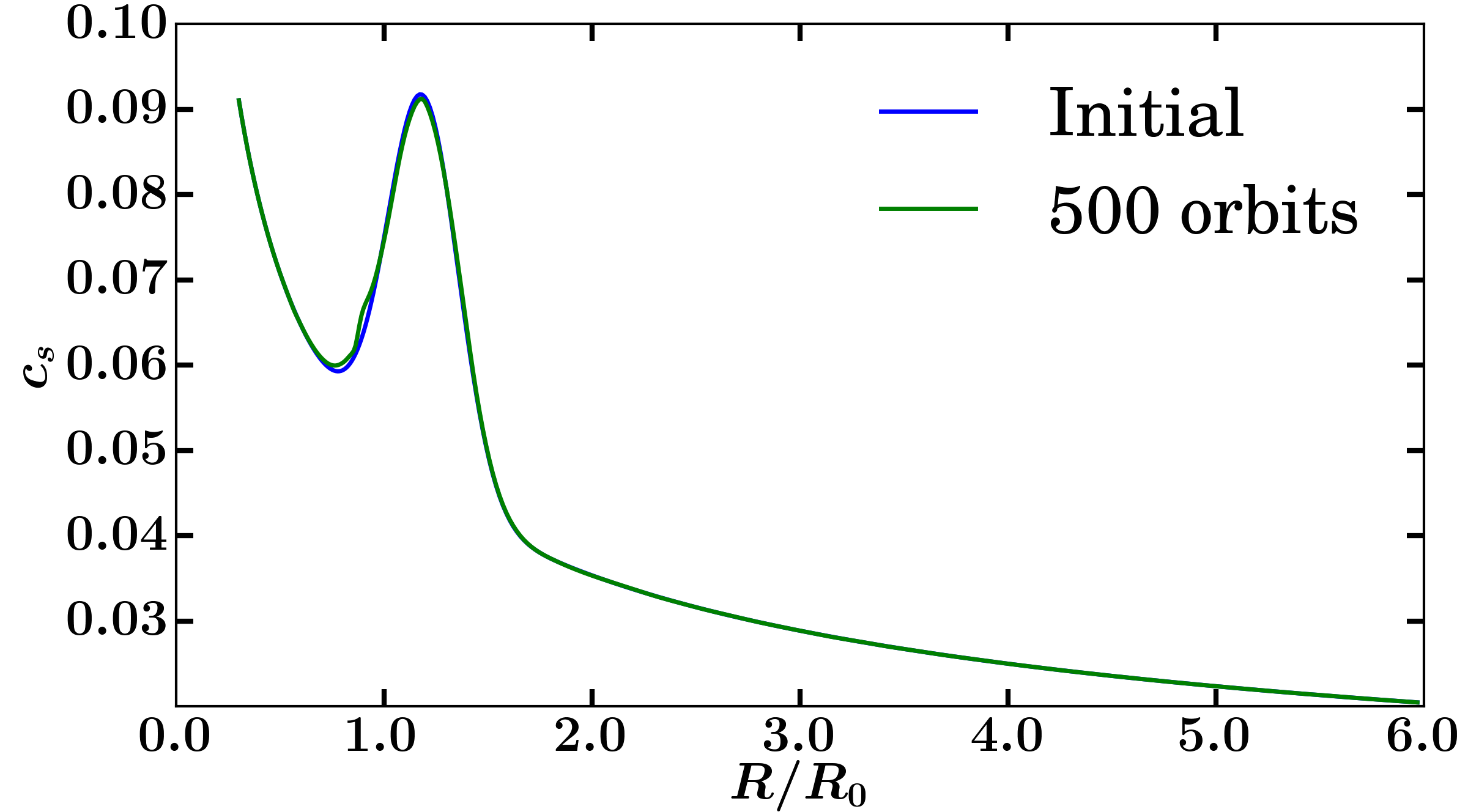}
	\vspace*{-5mm}
    \caption{An example showing two Gaussians in the sound speed distribution of a disc containing a planet, 500 orbits apart. One Gaussian is the initial Gaussian, while the other shows the same Gaussian after 500 orbits have elapsed. The disc setup is described in Section \ref{Sec:NumSet} and the Gaussian parameters are described in Section \ref{SubSec:Stand}. Here we can see that there is some change in the Gaussian across these 500 orbits, and while it is only a small difference, it is not unnoticable.}
    \label{Fig:Init_500_compare}
\end{figure}

\begin{figure}
	\includegraphics[width=\columnwidth]{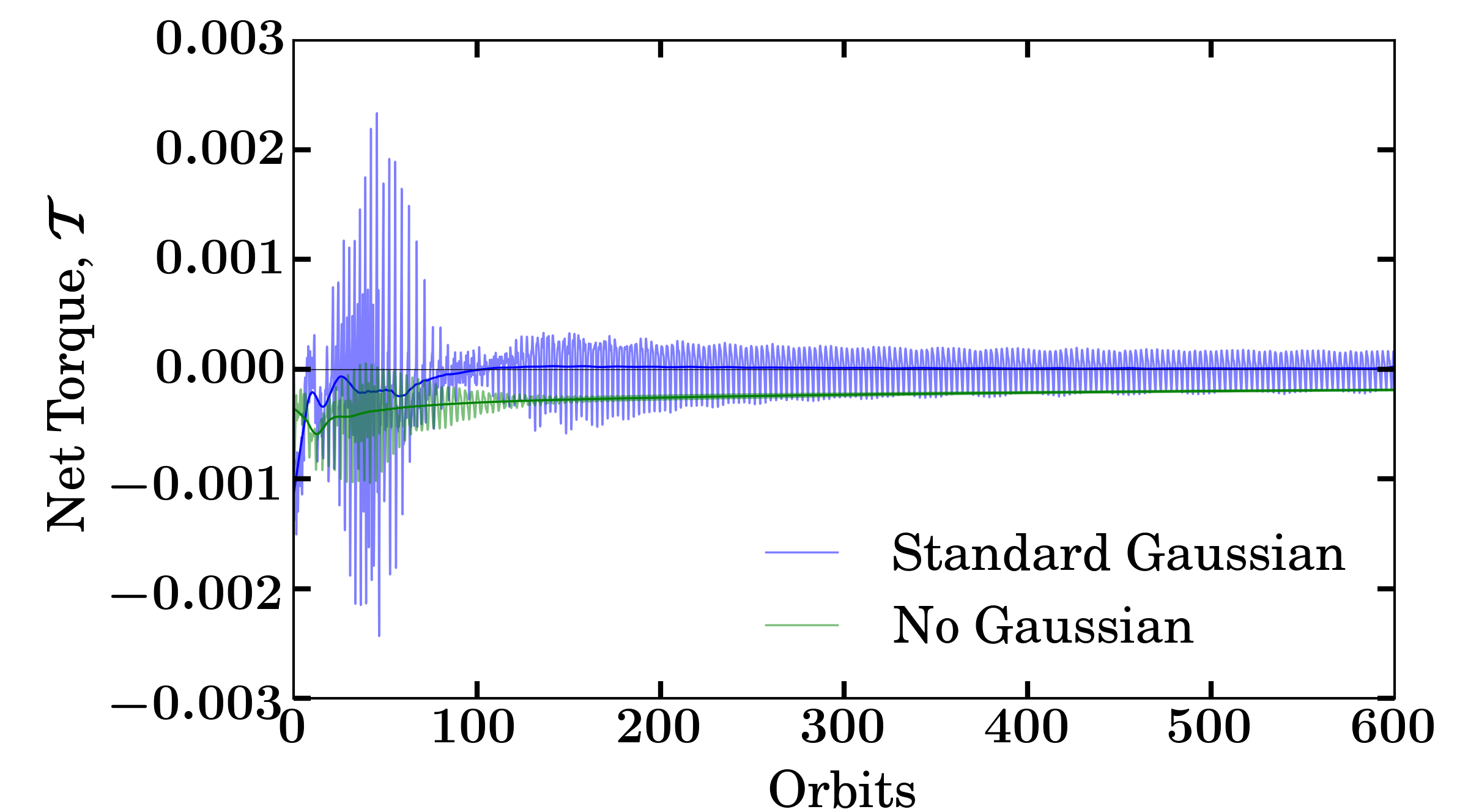}
	\vspace*{-5mm}
    \caption{Net torque on the planet for a disc containing a Gaussian in the sound speed profile and a disc containing no Gaussian in the sound speed profile. This is an example of how instabilities excited by the presence of the planet and Gaussian effect the net torque on the planet, for a Gaussian setup described in more detail in Section \ref{SubSec:Stand}. Clearly we can see the system reaches an approximate equilibrium, from which averaging over the last $100$ orbits provides us with a good description of the net torque. The bold lines represent the locally smoothed average of the net torque.}  
    \label{Fig:Torques}
\end{figure}

Even ensuring that the Gaussians themselves are stable in a disc containing no planet, we find that it is very difficult to stop them from going unstable once a planet has been introduced. Keeping our Gaussians in the Rossby stable regime is still important as it confirms that any instability present is the result of excitation by the planet, and not an intrinsic property of the Gaussian itself. The magnitude of the excited instability can greatly vary with the parameters of the Gaussian, similarly to the Rossby and Rayleigh instabilities, and in some cases, such as for very low $A$, the instability is almost non-existent or is not even excited. Unfortunately it seems impossible to completely remove these instabilities from our simulations. These instabilities, in addition to the presence of the planet, act to perturb the Gaussian profile such that the equilibrium profile is different from the initial profile. However, as long as the instabilities are weak the difference between these profiles is very small. This can be seen in Figure \ref{Fig:Init_500_compare}. Accepting that these instabilities will always be present, the problem now is do they impact the net torque on the planet? This is investigated in more depth when we discuss the variation of $\mathcal{W}$ in Section \ref{SubSec:Res_FWHM}, but we find that the average net torque appears to be unaffected by the presence of the instabilities. However, these instabilities do impact the amplitude of the net torque variation, meaning that particularly violent instabilities produce a much greater amplitude. This can increase the error in the net torque averaged over the last 100 orbits and can also impact the shape of the net torque distribution over these orbits, resulting in a state that is further from equilibrium than a more stable case. As a result we must be careful to not use Gaussians that excite violent instabilities, to preserve the validity of our average net torque result. An example of a result in which instabilities are excited but the net torque remains largely unchanged is given in Figure \ref{Fig:Torques}. Here we can see the variability in net torque is larger in the Gaussian case than in the no Gaussian case, however the locally smoothed average is not impacted by this variability. It is important to note that the strength of the instability decreases with time, such that many of the instabilities that are weak at $500$ orbits are much stronger at $100$ orbits.

Finally we decide to average the net torque from $500-600$ orbits as it is clear that the simulation has reached an approximate equilibrium. To ensure this is a valid assumption we have run a number of our simulations to $3000$ orbits (including that shown in Figure \ref{Fig:Torques}) in order to check the behaviour of the torque over large timescales and we find no significant deviation from the approximate equilibrium.

Between our findings that $500$ orbits is a good approximation to equilibrium and that the instabilities do not have a great impact on the net torque, we feel confident that averaging the net torque from $500 - 600$ orbits provides an accurate description of the overall net torque.

\section{Results}\label{Sec:Res}

\subsection{Reversing the Direction of Migration}\label{SubSec:Stand}
\begin{figure}
	\includegraphics[width=\columnwidth]{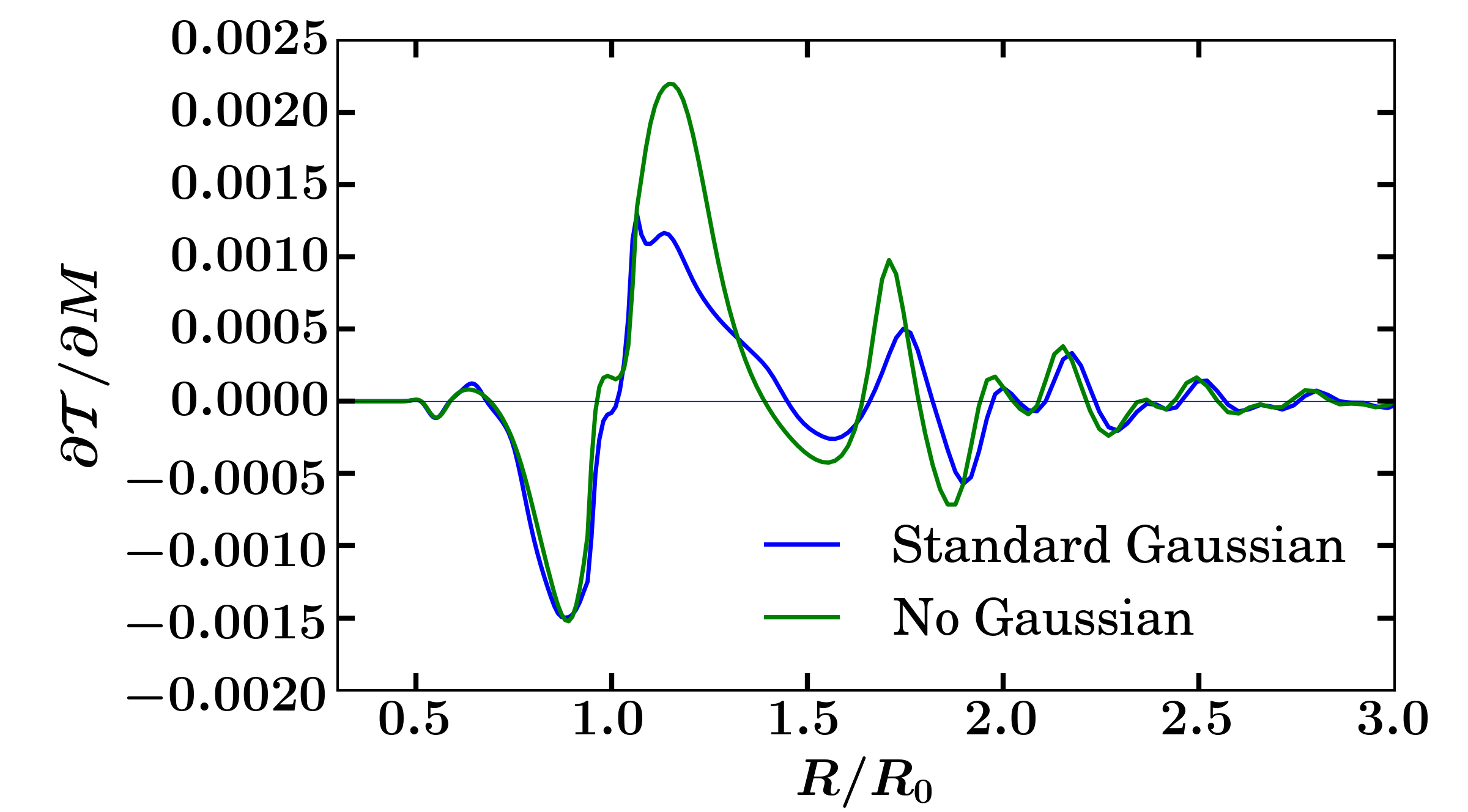}
	\vspace*{-5mm}
    \caption{Azimuthally averaged torque profiles for a disc containing a Gaussian in the sound speed profile and a disc containing no Gaussian in the sound speed profile. This Gaussians parameters are discussed in detail in Section \ref{SubSec:Stand}. Torque is calculated as described in Section \ref{SubSec:TorqueCalc} and averaged over $100$ orbits after the simulation reaches approximate equilibrium at orbit $500$. This clearly shows the reduction in the net torque contribution from the planets outer wake due to the irradiation of the outer gap edge.}
    \label{Fig:Torque_Profile}
\end{figure}

Using the setup described in Section \ref{Sec:NumSet} and results of Section \ref{Sec:Just} as a guideline for parameter space we investigated the net torque on a gap opening planet with a number of different Gaussians in the sound speed profile acting as stellar heating at the gap edge. We aimed to investigate whether this could change the sign of the net torque on the planet, as a negative net torque would result in inwards migration, and a positive net torque would result in outwards migration. We found that this method of heating can reduce the rate at which the planet migrates inwards, and can even change the sign of the net torque on the planet and produce outwards migration, as can be seen in Figure \ref{Fig:Torques}. Not only that, but we found the net torque on the planet can be positive even for the limited parameter space provided from Section \ref{Sec:Just}, in which the Gaussians are stable and the amplitude of the heating is in good agreement with the expected heating from a number of classifications of stars, assuming the gap edge is exposed to maximum or near maximum stellar irradiation. From this we settle on a standard case in which the direction of migration is reversed, for which $A = 2.0$, $\mathcal{W} = 0.40$ and $R_G = 1.20$. This returns a net torque of $\mathcal{T} = 7.395\times10^{-6}$, compared to the net torque for a planet and no Gaussian heating, $\mathcal{T} = -1.939\times10^{-4}$. This positive torque is two orders of magnitude smaller than the normal case for a planet and no Gaussian, hence we can deduce that although the direction of migration has been reversed, the rate of migration will be significantly slower. Figure \ref{Fig:Torque_Profile} clearly shows the impact of this Gaussian in reducing the net torque contribution from the planets outer wake. Using this setup as a standard case, we can now investigate the sensitivity of the net torque to the parameters of the Gaussian.

\subsection{Torque Dependence on $\mathcal{W}$}\label{SubSec:Res_FWHM}
The first parameter we investigate is $\mathcal{W}$. As this parameter is heavily constrained by the onset of instabilities for $\mathcal{W}<0.40$, this gives us little parameter space to work with. As a result we still investigate the impact on the torque for $\mathcal{W}<0.40$, but consider that they are unstable and as such the resultant Gaussian is very likely to be different to that which we describe. The reasoning is to get a larger picture of how $\mathcal{W}$ effects the net torque, however, outwards migration from unstable Gaussians will not be regarded as a viable solution to reversing the direction of migration.

The net torques for different $\mathcal{W}$ are listed in Table \ref{Table:FWHM}. From this we can see that net torque does not have a simple dependence on $\mathcal{W}$, rather the net torque peaks for a certain $\mathcal{W}$, and increasing $\mathcal{W}$ beyond this reduces the net torque. Fortunately for this setup the net torque remains positive for large $\mathcal{W}$, in which the Gaussians are stable, however it is clear that increasing $\mathcal{W}$ further will continue to reduce the net torque. This is likely due to the Gaussian now infringing on the inner wake excited by the planet, which also contributes to the net torque on the planet. In a similar regard having $\mathcal{W}$ too low also lowers the net torque on the planet, which is likely due to the Gaussian only acting on a small radial strip of the disc, meaning that a large portion of the torque profile is unaffected by its presence. It is also important to reiterate here that for unstable Gaussians, especially those in the Rayleigh and Rossby unstable regime, it is unclear how much the instability effects the net torque and as such that could be producing the interesting $\mathcal{W}$ torque dependence we see here. Despite this Table \ref{Table:FWHM} shows us that, at least in the Rayleigh stable regime, that increasing $\mathcal{W}$ lowers the value of the net torque, which is a very important conclusion as stable Gaussians demand a large $\mathcal{W}$. 

\begin{table}
\caption{Net Torque on the planet for stable/unstable Gaussians with different $\mathcal{W}$ at $A = 2.00$, $R_G = 1.20$}
\centering
\begin{tabular}{c|c|c}
\hline
$\mathcal{W}$ & $\mathcal{T}$ & Stability \\
\hline
0.10 & $1.653\times10^{-5}$ & Rayleigh and Rossby unstable \\
0.20 & $3.712\times10^{-5}$ & Rayleigh and Rossby unstable\\
0.30 & $4.115\times10^{-5}$ & Rossby unstable\\
0.35 & $2.585\times10^{-5}$ & Rossby unstable\\
0.40 & $7.395\times10^{-6}$ & Stable\\
0.45 & $-1.730\times10^{-5}$ & Stable \\
0.50 & $-3.942\times10^{-5}$ & Stable \\

\hline
\end{tabular}
\label{Table:FWHM}
\end{table}

\begin{figure}
	\includegraphics[width=\columnwidth]{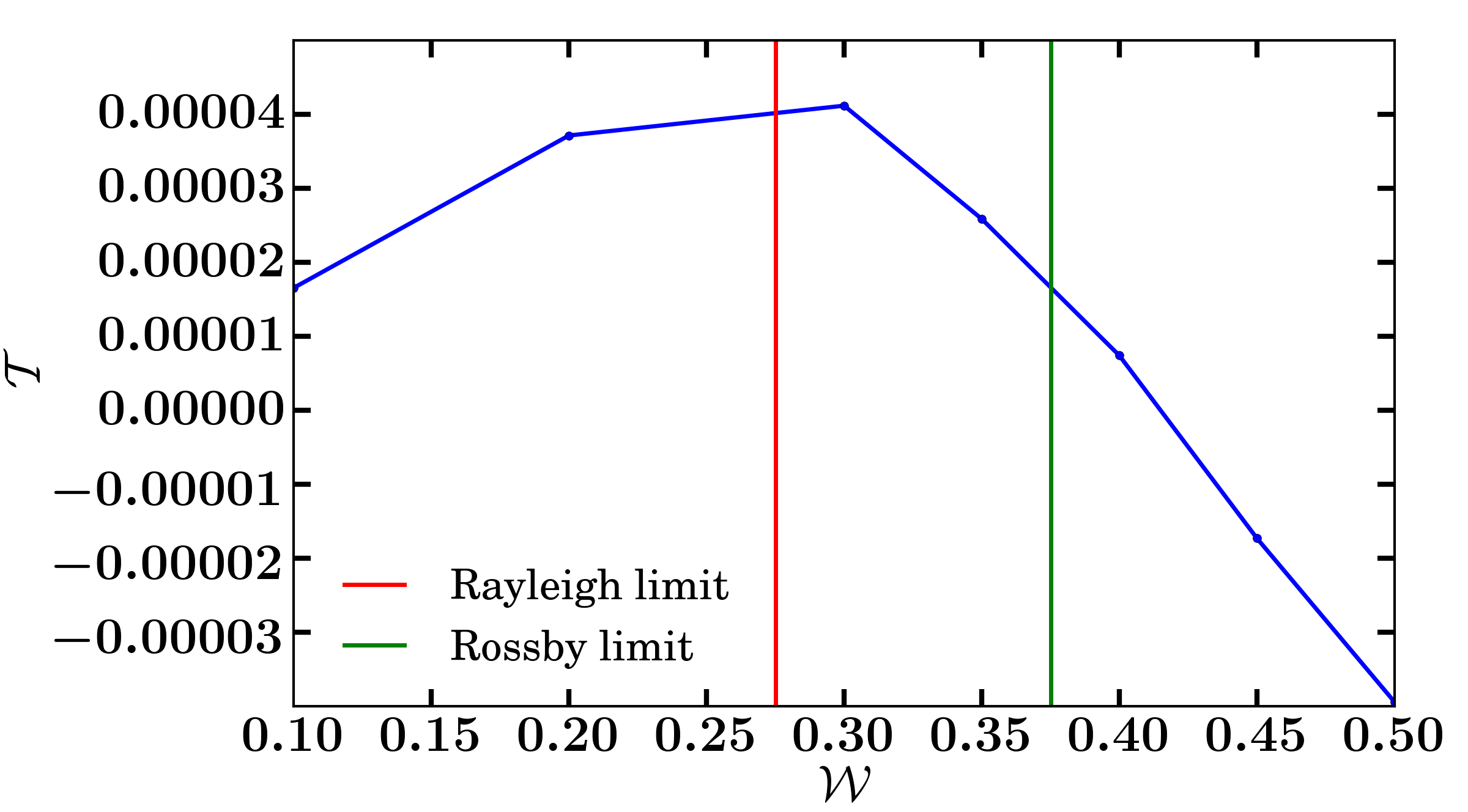}
	\vspace*{-5mm}
    \caption{The change in net torque on the planet as $\mathcal{W}$ is increased. Here we can see that the increase in $\mathcal{W}$ causes an increase in torque in the Rayleigh unstable regime and a decrease in torque in the Rayleigh stable regime. Also it is apparent that the excitement of the Rossby Wave instability does not influence the net torque on the planet as the trend is unaffected by the crossing of the Rossby stability limit. The Rossby and Rayleigh stability limits shown here are at an estimate of the $\mathcal{W}$ at which they occur, as we have only investigated their location to a $\mathcal{W}$ resolution of $0.05$.}
    \label{Fig:TqFWHM}
\end{figure}

Figure \ref{Fig:TqFWHM} also shows the results presented in Table \ref{Table:FWHM}. Here it is easier to see the impact of changing $\mathcal{W}$ on the net torque and also the behaviour of the torque as it crosses the regions of Rayleigh and Rossby stability. From this we can see that the torque increases with $\mathcal{W}$ in the Rayleigh unstable regime, but outside this regime it falls with increasing $\mathcal{W}$. The most important thing to note from Figure \ref{Fig:TqFWHM} is that the trend in the net torque does not change as $\mathcal{W}$ crosses the Rossby stability limit, hence leading us to believe that the Rossby Wave instability does not influence the net torque on the planet.

\subsection{Torque Dependence on $\mathbf{A}$}
We investigate modifying the parameter $A$ for the range $A = 1.5 - 2.5$, consistent with our findings from Section \ref{Sec:Just}. Here the value of $A$ could represent the heating of the gap edge at maximum irradiation from the star or a reduced irradiation from a star that at maximum would give a higher $A$ value. Reasonable $A$ values for a variety of stellar classifications are given in Table \ref{Table:Stars}.  However we already know from Sections \ref{Sec:Just} and \ref{Sec:Inst} that $\mathcal{W}$ is closely linked to the $A$ of the Gaussian, to ensure we do not enter Rossby or Rayleigh unstable regimes. Therefore in a deviation from modifying only one parameter, we must also increase $\mathcal{W}$ to the appropriate value for stability for each value of $A$. These $\mathcal{W}$ are given in Table \ref{Table:Rossby}.
 
We find that modifying the $\mathcal{W}$ and $A$ of the Gaussian each have the opposite effect on the net torque on the planet. Section \ref{SubSec:Res_FWHM} already shows that outside of the Rayleigh unstable regime increasing $\mathcal{W}$ acts to move the net torque on the planet towards the negative. We find that increasing $A$ acts to move the net torque on the planet towards the positive and hence the effect of modifying both of these parameters compete with each other. However, $\mathcal{W}$ must be increased so much for larger values of $A$ that the effect of $\mathcal{W}$ on the net torque is stronger than the effect of the parameter $A$. This results in stable Gaussians with a larger $A$ actually making the net torque on the planet more negative and being unable to reverse the direction of migration. We then find that, at $R_G = 1.20$, there is a "sweet spot" where the effect of increasing $A$ on the torque overcomes the effect of increasing $\mathcal{W}$, and we get outwards migration. This information is presented in Table \ref{Table:A}, which shows that in the range $1.75 \leq A \leq 2.15$ we can return outwards migration for stable Gaussians.

\begin{table}
\caption{Net Torque on the planet for stable Gaussians with different values of $A$ at $R_G = 1.20$}
\centering
\begin{tabular}{c|c|c}
\hline
$A$ & $\mathcal{W}$ & $\mathcal{T}$ \\
\hline
1.50 & 0.25 & $-2.742\times10^{-5}$ \\
1.75 & 0.35 & $1.022\times10^{-6}$ \\
2.00 & 0.40 & $7.395\times10^{-6}$ \\
2.15 & 0.45 & $2.775\times10^{-6}$ \\
2.25 & 0.50 & $-1.149\times10^{-5}$ \\
2.50 & 0.60 & $-2.499\times10^{-5}$ \\

\hline
\end{tabular}
\label{Table:A}
\end{table}

\subsection{Torque Dependence on $\mathbf{R_{G}}$}
So far all of the variations made to parameters of the standard Gaussian have held the location of the Gaussian constant, $R_G  = 1.20$. We now move the Gaussian radially outwards across the disc to $R_G = 1.30 - 1.40$, still located on the gap edge but corresponding to a lower opacity disc, as radiation from the star now penetrates further into the gap edge. The immediate change that results from moving the Gaussian is that the $\mathcal{W}$ required to stabilise the Gaussian increases. This is because the Gaussian now resides at a location with a larger aspect ratio. Therefore we can no longer use the results presented in Table \ref{Table:Rossby} and instead recalculate the region for which the Gaussians are stable. This information is presented in Table \ref{Table:Rossby2}.

\begin{table}
\caption{$\mathcal{W}$ limits for Rossby and Rayleigh stable Gaussians of height $A$}
\centering
\begin{tabular}{c|c|c|c}
\hline
	&& Rayleigh Stable & Rossby and Rayleigh Stable \\
$R_G$ &$A$ & $\mathcal{W}$ & $\mathcal{W}$ \\
\hline
1.30 & 1.50 & 0.20 & 0.30 \\
1.30 & 1.75 & 0.30 & 0.35 \\
1.30 & 2.00 & 0.35 & 0.45 \\
1.30 & 2.15 & 0.35 & 0.50 \\
1.30 & 2.25 & 0.40 & 0.55 \\
1.30 & 2.50 & 0.45 & 0.65 \\
1.35 & 1.50 & 0.20 & 0.30 \\
1.35 & 1.75 & 0.30 & 0.35 \\
1.35 & 2.00 & 0.35 & 0.45 \\
1.40 & 1.50 & 0.25 & 0.30 \\
1.40 & 1.75 & 0.30 & 0.40 \\
1.40 & 2.00 & 0.35 & 0.45 \\
1.40 & 2.25 & 0.40 & 0.55 \\
\hline
\end{tabular}
\label{Table:Rossby2}
\end{table}

\begin{figure}
	\includegraphics[width=\columnwidth]{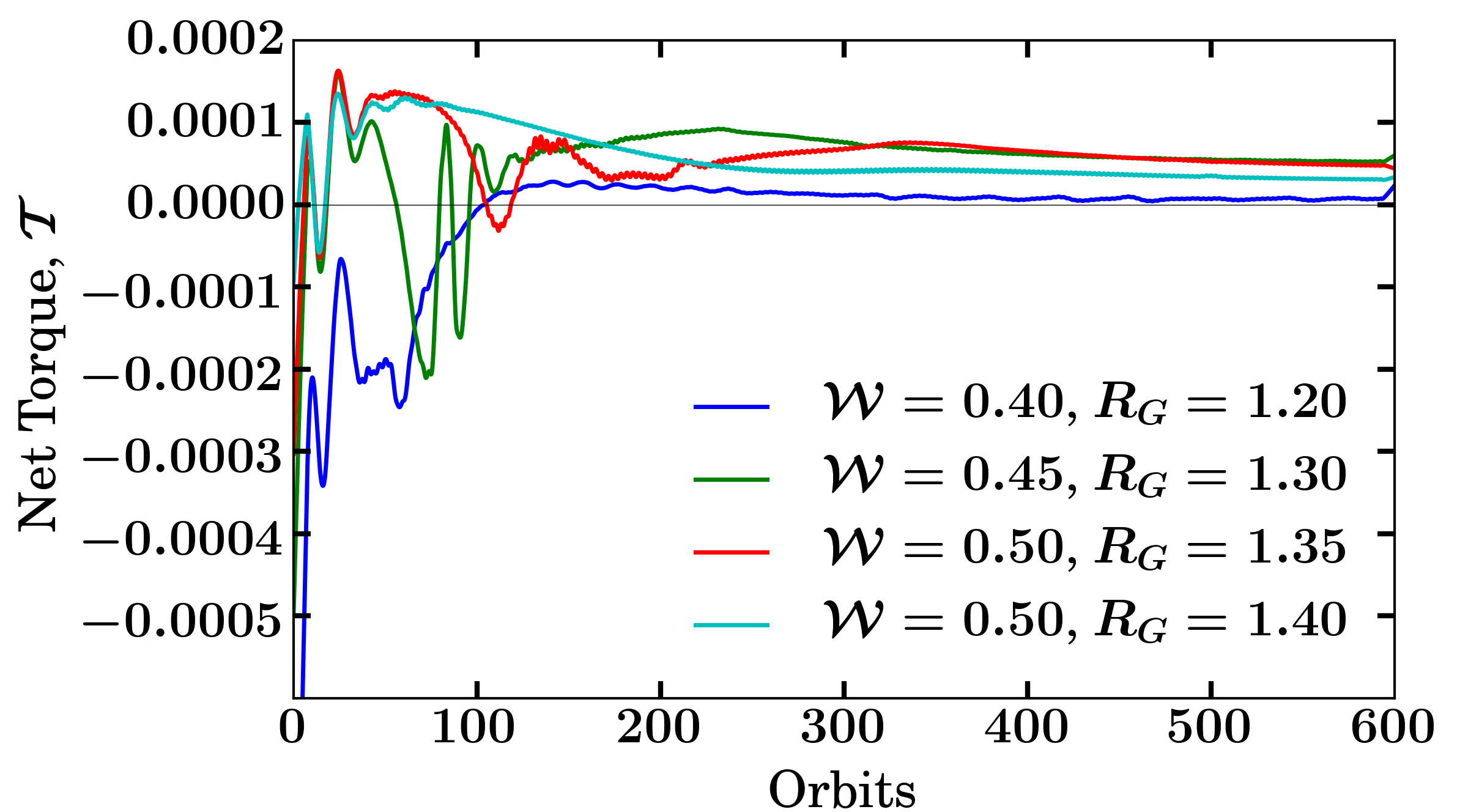}
	\vspace*{-5mm}
    \caption{Smoothed net torque curves for a number of the simulations presented in Table \ref{Table:A2}. This shows the improvement made in returning a positive net torque when we increase the radius at which the Gaussian in the sound speed profile is centred. This increased net torque corresponds to more parameter space for outward migration at these larger radii.} 
    \label{Fig:Multi_Torques}
\end{figure}

A concern with further increasing $\mathcal{W}$ was outlined in Section \ref{Sec:Just}, namely that the wider the Gaussian the more the heating infringes onto the gap region and subsequently could reduce the torque contribution from the inner wake. In this case it appears that the increase in $\mathcal{W}$ is significantly smaller than the distance the Gaussian was moved, $\Delta R_G = 0.10 - 0.20$, hence the inner wake will be unaffected by the increase in $\mathcal{W}$. We now use these $\mathcal{W}$ as the limit for stable Gaussians, similarly to the way we used $\mathcal{W}$ in Table \ref{Table:Rossby} as the limit for stable Gaussians at $R_G = 1.20$. With this in mind we replicate Table \ref{Table:A} but for stable Gaussians at $R_G = 1.30 - 1.40$. This information is presented in Table \ref{Table:A2}. We also present a number of these results in Figure \ref{Fig:Multi_Torques}, which shows the smoothed net torque for each simulation, similar to that shown in Figure \ref{Fig:Torques}. The deviation in the last $<10$ orbits in each curve is an artefact of the smoothing mechanism we use, and should be disregarded.

\begin{table}
\caption{Net Torque on the planet for stable Gaussians with different input parameters}
\centering
\begin{tabular}{c|c|c|c}
\hline
$R_G$ & $A$ & $\mathcal{W}$ & $\mathcal{T}$ \\
\hline
1.30 & 1.50 & 0.30 & $-4.395\times10^{-5}$ \\
1.30 & 1.75 & 0.35 & $3.123\times10^{-5}$ \\
1.30 & 1.75 & 0.40 & $2.432\times10^{-5}$ \\
1.30 & 2.00 & 0.45 & $5.270\times10^{-5}$ \\
1.30 & 2.00 & 0.50 & $3.431\times10^{-5}$ \\
1.30 & 2.15 & 0.50 & $4.845\times10^{-5}$ \\
1.30 & 2.25 & 0.55 & $3.378\times10^{-5}$ \\
1.30 & 2.50 & 0.65 & $1.449\times10^{-5}$ \\
1.35 & 1.50 & 0.30 & $-9.006\times10^{-5}$ \\
1.35 & 1.75 & 0.35 & $4.177\times10^{-5}$ \\
1.35 & 1.75 & 0.40 & $-4.802\times10^{-7}$ \\
1.35 & 2.00 & 0.45 & $6.710\times10^{-5}$ \\
1.35 & 2.00 & 0.50 & $4.908\times10^{-5}$ \\
1.40 & 1.50 & 0.30 & $-1.323\times10^{-4}$ \\
1.40 & 1.75 & 0.40 & $-4.661\times10^{-5}$ \\
1.40 & 2.00 & 0.45 & $6.008\times10^{-5}$ \\
1.40 & 2.00 & 0.50 & $3.178\times10^{-5}$ \\
1.40 & 2.25 & 0.55 & $3.870\times10^{-5}$ \\
1.40 & 2.25 & 0.60 & $6.140\times10^{-5}$ \\

\hline
\end{tabular}
\label{Table:A2}
\end{table}

The results shown in Table \ref{Table:A2} are very interesting as they show that there is much larger potential for outward migration for heating at these larger radii. The net torque at these locations is also often an order of magnitude larger than at $R_G = 1.20$, meaning that the error in the net torque calculation will be small compared to the actual value and hence we can be confident in our result. This increase in magnitude of net torque also means that outward migration will be more rapid for heating at this location, or in the case of a negative torque, inwards migration will be slower.

We briefly investigated modifying $\mathcal{W}$ while keeping $A$ and $R_G$ constant, in a similar manner to Section \ref{SubSec:Res_FWHM}. It was found that the net torque behaves in the same way as it does at $R_G = 1.20$ and increasing the $\mathcal{W}$ of the Gaussian too much begins to reduce the net torque. This was not unexpected, as only the location of the Gaussian has changed, but the overall more positive net torque at $R_G = 1.30 - 1.40$ means that $\mathcal{W}$ can be increased more than it can at $R_G = 1.20$ while still retaining outwards migration. This is another example of the increased parameter space available at this location for stable Gaussians with outward migration, which gives us greater confidence in saying that outwards migration could potentially occur here. 

\subsection{Exploring Asymmetric Distributions}\label{SubSec:Res_Asym}

As discussed in Section \ref{Sec:Just} we intend our Gaussian distribution to only reduce the net torque contribution from the planet's outer wake. From our results in Section \ref{SubSec:Res_FWHM} we can see that should $\mathcal{W}$ be increased too much the net torque from the planet's inner wake is also reduced. Therefore, in order to investigate the effect of increasing $\mathcal{W}$ without allowing the Gaussian to further reduce the torque contribution from the planets inner wake, we modify our description of the Gaussian. We name our new distribution an ``Asymmetric Gaussian distribution'' as our heating method remains based around a Gaussian of height $A$ at radius $R_G$, however we now have two different $\mathcal{W}$'s, one for the inner half of the Gaussian ($\mathcal{W_\textrm{in}}$, heating the gap edge and gap region) and one for the outer half of the Gaussian ($\mathcal{W_\textrm{out}}$, heating the gap edge and outer disc). Therefore we can now hold $\mathcal{W_\textrm{in}}$ constant at the smallest stable Gaussian $\mathcal{W}$ while extending $\mathcal{W_\textrm{out}}$ further into the outer disc. The results of running a number of asymmetric Gaussians with different initial parameters are presented in Table \ref{Table:Asymmetric}.

From Table \ref{Table:Asymmetric} we can see there is minimal difference in the net torque when we increase $\mathcal{W_\textrm{out}}$. This is especially true for the largest radii results, at $R_G = 1.40$, however for the smallest radii $R_G = 1.20$ the net torque increases by a significant amount. This shows that at $R_G = 1.20$ the net torque contribution from the outer wake is not fully covered by the Gaussian, whereas at larger radii the Gaussian already covers the wake, and the Gaussian now encroaches on the less perturbed outer disc region. This result further increases the parameter space  for which outward migration is allowed, as although torque is often reduced by extending $\mathcal{W_\textrm{out}}$ there is still room to extend $\mathcal{W_\textrm{out}}$ before the torque becomes negative again. Also extending $\mathcal{W_\textrm{out}}$ appears to increase the net torque at $R_G = 1.20$, which could potentially solve the problem of the limited parameter space for outward migration at this location.

\begin{table}
\caption{Net Torque on the planet for stable asymmetric Gaussians with different input parameters.}
\centering
\begin{tabular}{c|c|c|c|c}
\hline
$R_G$ & $A$ & $\mathcal{W_\textrm{in}}$ & $\mathcal{W_\textrm{out}}$ &$\mathcal{T}$ \\
\hline
1.20 & 1.75 & 0.35 & 0.35 & $1.022\times10^{-6}$ \\
1.20 & 1.75 & 0.35 & 0.45 & $3.527\times10^{-6}$ \\
1.20 & 2.00 & 0.40 & 0.40 & $7.395\times10^{-6}$ \\
1.20 & 2.00 & 0.40 & 0.50 & $1.205\times10^{-5}$ \\
1.20 & 2.00 & 0.40 & 0.60 & $1.430\times10^{-5}$ \\
1.30 & 1.75 & 0.35 & 0.35 & $3.123\times10^{-5}$ \\
1.30 & 1.75 & 0.35 & 0.45 & $2.227\times10^{-5}$ \\
1.30 & 1.75 & 0.40 & 0.40 & $2.432\times10^{-5}$ \\
1.30 & 1.75 & 0.40 & 0.50 & $1.488\times10^{-5}$ \\
1.30 & 2.00 & 0.45 & 0.45 & $5.270\times10^{-5}$ \\
1.30 & 2.00 & 0.45 & 0.55 & $5.147\times10^{-5}$ \\
1.30 & 2.00 & 0.45 & 0.65 & $5.270\times10^{-5}$ \\
1.30 & 2.00 & 0.50 & 0.50 & $3.431\times10^{-5}$ \\
1.30 & 2.00 & 0.50 & 0.60 & $3.471\times10^{-5}$ \\
1.30 & 2.00 & 0.50 & 0.70 & $3.506\times10^{-5}$ \\
1.35 & 1.75 & 0.35 & 0.35 & $4.177\times10^{-5}$ \\
1.35 & 1.75 & 0.35 & 0.45 & $-1.412\times10^{-5}$ \\
1.35 & 1.75 & 0.40 & 0.40 & $-4.802\times10^{-7}$ \\
1.35 & 1.75 & 0.40 & 0.50 & $-6.146\times10^{-6}$ \\
1.35 & 2.00 & 0.45 & 0.45 & $6.710\times10^{-5}$ \\
1.35 & 2.00 & 0.45 & 0.55 & $6.552\times10^{-5}$ \\
1.35 & 2.00 & 0.50 & 0.50 & $4.908\times10^{-5}$ \\
1.35 & 2.00 & 0.50 & 0.60 & $5.323\times10^{-5}$ \\
1.40 & 1.75 & 0.40 & 0.40 & $-4.661\times10^{-5}$ \\
1.40 & 1.75 & 0.40 & 0.50 & $-4.709\times10^{-5}$ \\
1.40 & 2.00 & 0.45 & 0.45 & $6.008\times10^{-5}$ \\
1.40 & 2.00 & 0.45 & 0.55 & $2.183\times10^{-5}$ \\
1.40 & 2.00 & 0.50 & 0.50 & $3.178\times10^{-5}$ \\
1.40 & 2.00 & 0.50 & 0.60 & $2.416\times10^{-5}$ \\

\hline
\end{tabular}
\label{Table:Asymmetric}
\end{table}

\section{Discussion}\label{Sec:Disc}
In this paper we have found that it is possible to reduce the magnitude of a planets inwards Type II migration rate by heating the outer gap edge due to stellar irradiation.  We have also found for a range of parameters that it is possible to reverse the direction of migration of the planet by causing the disc to exert a positive net torque on the planet, if the gap edge irradiation is the maximum or near maximum possible from the star. We represent the irradiation of the outer gap edge by initialising a Gaussian distribution in the sound speed profile of the disc at the gap edge, which reduces the contribution of the planets outer wake on the net torque on the planet. We then measure the resultant net torque on the planet and use this to infer the relative rate and direction of migration.

We make a number of assumptions and approximations in our simulations which allow us to investigate a simplified model of the system we describe. Primarily, we are investigating a three dimensional problem using two dimensional simulations. This is a result of the increased computational time required for three dimensional simulations meaning it would be impossible to perform a parameter study of this magnitude within a reasonable time frame. Therefore we present these results as an exploration into the effect of heating the outer gap edge and should be followed up in three dimensions. We also use a very simplified model for the heating of the gap edge, namely an increase in the sound speed profile of the disc at this location following a Gaussian distribution. Realistically the three dimensional location and radial dependence of this heating would be a function of the disc's opacity, which is not considered here. The magnitude of our heating of the outer gap edge is also calculated ignoring the presence of the planet and the inner disc (see Section \ref{Sec:Just}). Therefore we present our results at maximum to near maximum possible gap edge irradiation from the star, however the inner disc and planet should be accounted for in more realistic estimates of the gap edge heating. As such, the heating of the outer gap edge should also be explored numerically using three dimensional radiative transfer simulations.

An important concept in our work is the reduction of the net torque contribution from the planets outer wake, while leaving the inner wake unchanged. This is not necessarily always the case in our simulations. We expect some torque difference at the inner wake due to the presence of the Gaussian in the sound speed profile, an unfortunate consequence of the required $\mathcal{W}$ to avoid the Gaussians becoming inherently unstable. However, we find even in cases in which $\mathcal{W_\textrm{in}}$ of an asymmetric Gaussian is held constant and $\mathcal{W_\textrm{out}}$ is modified the torque contribution from the inner wake can still change, albeit a small amount. This is because the torque does not depend solely on the sound speed, it also has a surface density dependence. As the gap profile is modified by the change in sound speed, this can have an impact on the torque across the gap region. It is also known that the gap edge steepness, gap width and gap depth can all impact the torque \citep{Petrovich2012}. Overall, the change in the torque at the inner wake is negligible, especially compared to the change in the torque at the outer wake, therefore this is largely irrelevant to our results. 

The selected $\mathcal{W}$'s of our Gaussians is also an important point for discussion in our work. The reasons for selecting the values we do has already been discussed in Sections \ref{Sec:Just} and \ref{Sec:Inst}, however here we shall briefly mention further difficulties surrounding our selection of $\mathcal{W}$. The $\mathcal{W}$ we have selected are entirely based off of stability arguments from Section \ref{Sec:Inst}. Hence we keep our $\mathcal{W}$ close to the stability cut off. However, $\mathcal{W}$ realistically would represent a mixture of the absorption of stellar radiation by the disc at locations outside the peak in absorption efficiency ($R_G$) and the self heating of the disc by re-emission of radiation from the location of maximum absorption efficiency. This means that $\mathcal{W}$ would be closely linked to the opacity of the disc and its rate of heat transport. By selecting a $\mathcal{W}$, and an $R_G$, we are indirectly making statements about these parameters. Of course these would depend on more than just these Gaussian parameters, such as the composition of the disc and the gas to dust ratio, however these are non-trivial to gain exact values for and so we do not directly consider them in this work. Taking these into account will likely modify the range of parameter space for which the Gaussian is realistic heating for a disc, and therefore further constrain the outwards migration regime.

We make further simplifying assumptions within the model we have outlined above, which we do not believe will affect the overall conclusion, that outwards migration is possible, however will undoubtedly affect the exact numerical results. We use a non flaring disc, while claiming that, theoretically, the flaring of the disc is what allows the stellar radiation to impact the gap edge and thus the heating to occur. This is a simplifying assumption that may affect the structure of the outer disc and the heat transport outwards, but will largely not affect the net torque on the planet.

We consider a simplified prediction of the magnitude of heating from a number of stellar sources provided in Table \ref{Table:Stars}. These predictions agree with the $A$ parameters we use within our simulations, with parameter space large enough that should a more complex heating model provide slightly different heating magnitudes (most likely lower) we would still expect outwards migration in a number of cases. As stated above we neglect the presence of the inner disc and planet when determining the magnitude of heating of the gap edge, such that we can follow the example of \cite{Dullemond2001}. Realistically the inner disc would block some of the radiation from the star, and hence this would likely further reduce the magnitude of heating. This effect would be difficult to quantify in this simple approximation, as the planet impacts both the inner and outer disc structure, hence radiative transfer simulations would be required to adequately investigate. In addition, it has been shown that reflection, absorption and re-radiation from the outer gap edge can cause the inner gap edge to also become heated, however to a lesser extent \citep{Turner2012}. This can cause a reduction in the one sided net torque at this location, and a "puffing up" of the inner gap edge, blocking more of the stellar radiation that otherwise would be incident on the outer gap edge. If the heating is significant, this could effect our results by causing the net torque on the planet to become more negative.

We address the usage of a Gaussian for the shape of the heating distribution in Section \ref{SubSec:Res_Asym}, and find that the result changes minimally in most cases for an asymmetric heating distribution. In fact in some cases we find this improves the likelihood of outwards migration, rather than reducing the net torque on the planet. This shows that our results are not specific to Gaussians distributions.

Our Gaussian heating setup does not capture the effect of shadowing the outer disc by the "puffed up" heated gap edge. This would reduce the temperature in the outer disc and as a result move the net torque on the planet towards the negative. As has been shown in Section \ref{SubSec:Res_FWHM} we require a few scale heights of heating to reduce the torque contribution from the planets outer wake, so if the cooling effect from self-shadowing does not allow for this then this could cause the net torque on the planet to become more negative. The results of \cite{Jang-Condell2013} imply that gap edge heating may occur over a region a few scale heights wide.

We heat the location of the gap edge before the gap has formed, as our planet is stationary and we know the location at which the gap will occur. This may cause non-physical effects at early times, as the gap edge cannot be visible to the star if it has not formed yet, but we are confident that by $500$ orbits the simulation has reached equilibrium and the result is independent of early time effects.

We use only stable Gaussians in our simulations, however it is clear that the presence of the planet excites instabilities in the Gaussian which is unavoidable. We are confident the instabilities have no effect on the net torque once we reach an approximate equilibrium, as while there is short term variability in the net torque on the planet the magnitude is approximately constant such that the average over $100$ orbits is a good indicator of the overall net torque on the planet. We are also confident that the Rossby Wave instability has no effect on the net torque on the planet, as shown in Figure \ref{Fig:TqFWHM}. Despite arguing that these instabilities have little to no effect on the overall net torque and that instabilities are unavoidable in simulations containing planets and Gaussians, we endeavour to avoid inputting instabilities into our simulations where we can control their excitement. Hence, we do not thoroughly investigate the net torque on the planet for the regime in which the input Gaussian is Rossby unstable, despite Section \ref{SubSec:Res_FWHM} showing us that this region can return outwards migration. From our results, it is not unbelievable that should a system similar to those we describe exist in nature, it could be unstable. In this case Rossby instabilities excited due to the gap edge irradiation could potentially occur and so this region would be important. From our results in Section \ref{SubSec:Res_FWHM} we believe this would only further open up parameter space for which outward migration is a possibility.

As the planet is stationary in the disc and our measure of migration comes from the net torque on the planet, the gap edge and therefore the Gaussian representing heating is also stationary. Realistically in Type II migration the gap edge would move with the planet as it migrates, and this would change the magnitude of heating on the gap edge. Hence the predicted net torque on the planet will not be constant as the planet moves and the shape of the input Gaussian should change as the gap edge moves. This means it is very possible that a planet will undergo both inwards and outwards migration as the system evolves, however investigating this is beyond the scope of this paper. 

We use a fixed value for the thermal relaxation timescale, $\tau = 2\pi$, in our simulations. This acts as a compromise between an adiabatic and an isothermal simulation, for which $\tau = 0$. This is because a longer thermal relaxation timescale will cause the simulation to damp back slower to the initial Gaussian in the sound speed profile. Hence the Gaussian heating acting on the disc will be less accurate to the initial distribution for larger $\tau$. \cite{Richert2015} have shown that for inefficient cooling timescales a planet in a two dimensional adiabatic disc can excite instabilities that significantly alter the disc structure. In some cases it is found that a massive planet will not even open a gap. Therefore, we select our thermal relaxation timescale to be much shorter than the gap formation timescale. Despite this, during our investigation we briefly experimented with increasing the thermal relaxation timescale, for the standard case of a stable Gaussian we describe in Section \ref{SubSec:Stand}. We found that increasing the thermal relaxation timescale to $\tau = 10\pi$ and $\tau = 20\pi$ greatly deformed the shape of the Gaussian and excited instabilities in the disc similar to those described in Section \ref{Sec:Inst}, even before a planet was introduced. From this and the results of \cite{Richert2015}, we decided that the thermal relaxation timescale was not a parameter to be investigated in our parameter study. It should be made clear though that in a realistic disc the local thermal relaxation timescale will vary with radius as, for example, in an optically thin outer disc heat will be redistributed faster than for an optically thick inner disc. As such, the results we present are valid for a thermal relaxation timescale of $\tau = 2\pi$ at the location of heating. Far from the location of heating the sound speed profile of the disc is unperturbed and as a result the shape of the profile is independent of the thermal relaxation timescale here, regardless of its realistic value. 

As we hold our planets on a fixed circular orbit there is no accounting for the eccentricity evolution of the planet's orbit. It is known that the eccentricity of a giant planet is excited by it's Lindblad resonances and damped by it's non-co-orbital corotation torques. Unless the gap formed is very large, the damping via corotation torques dominates and the eccentricity is not excited. However, it has been shown that the illumination of the outer gap edge by stellar irradiation can modify the planet's corotation torque to allow the eccentricity excitation via Lindblad resonances to dominate \citep{Tsang2014,Tsangb2014}. Hence this could change the planets orbital path from the circular case we have investigated, which could have implications on the net torque on the planet.

Our results act as proof that this method of modifying the classical Type II planetary migration may effect the migration of giant planets. Therefore, this method of reducing or reversing migration rate is important to consider when studying the evolution of planetary systems, either as a sole method of modifying planetary migration, or in tandem with other previously proposed methods, i.e. \cite{Coleman&Nelson2016}.  

\section{Summary and Conclusions}\label{Sec:Conc}
We have investigated the heating of the outer gap edge by stellar radiation using a Gaussian distribution in the sound speed profile of a disc. Using this method it is possible to lower the contribution to the net torque on the planet from the planets outer wake, while leaving the contribution from the inner wake mostly unchanged. The goal was to address the problem that observational evidence dictates that giant planets are more common at orbits $R>1\textrm{AU}$ \citep{Mayor2011,Cassan2012,Fressin2013,Santerne2016} and that for a gas giant planet to survive throughout the discs lifetime it must begin Type II migration at an orbital radius $R>20\textrm{AU}$ \citep{Coleman&Nelson2014}. At this radius the core accretion timescale exceeds the lifetime of the disc, therefore there must be a process that limits Type II migration speed \citep{Nelson2000}. We found heating the gap edge via stellar radiation is a method of reducing the net torque on the planet and by extension the rate of inwards migration. We also found that for the extreme case of maximum to near maximum gap edge irradiation it is possible to return a positive net torque on the planet, which corresponds to outwards migration for the planet. We have investigated the impact on the net torque of modifying the following Gaussian parameters:

\begin{itemize}
	\item $A$, the amplitude of the Gaussian.
	\item $\mathcal{W}$, the full width at half maximum of the Gaussian.
	\item $R_G$, the radial location of the Gaussian peak.
	\item The impact of asymetric Gaussians.
\end{itemize}

We found that the range of applicable Gaussians was limited by $\mathcal{W}$ constraints to avoid Rossby and Rayleigh unstable regimes. Our test case of $R_G = 1.20$ was studied extensively, and was found to provide very low magnitude positive torque for a small range of $A$ and $\mathcal{W}$, providing the potential for weak outward migration, or a large reduction in the rate of inwards migration. At larger radii, $R_G = 1.30 -1.40$, we found that there is a significantly larger parameter space for which outward migration is possible, both in $A$ and $\mathcal{W}$ and that the magnitude of the net torque is larger, so the planet is further from the edge of the outwards migration regime and has a higher rate of migration. We found that in general the height of the Gaussian must be at least $A = 1.75$ to achieve outwards migration and that increasing the $\mathcal{W}$ of the Gaussian lowers the net torque on the planet, a result of the Gaussian impacting the torque contribution from both the inner and outer wakes, rather than just the outer wake. We find that for a number of sample stellar classifications the predicted magnitude of heating at the gap edge corresponds to $A \approx 2.0$ at maximum or near maximum stellar irradiation with a weak radial dependence, $A \propto R_G^{1/8}$, which would make outwards migration a possibility. For more modest $A$, which could account for the presence of an inner disc, re-radiation heating the inner gap edge and self-shadowing of the outer disc, we still find significant effects on the net torque on the planet, meaning that even if outwards migration is not a possibility there is still the potential for reduction in Type II migration rate. We also investigate asymmetric Gaussian distributions, in which $\mathcal{W}$ is greater on the outer Gaussian edge, and at the limit of stability on the inner Gaussian edge. We found that increasing $\mathcal{W_\textrm{out}}$ can both increase the net torque on the planet at smaller radii ($R_G = 1.20$) and decrease the net torque at larger radii, such as the reduction at $R_G = 1.40$. This could potentially solve the problem of the small parameter space for outwards migration at $R_G = 1.20$, while the weak reduction in net torque for ranges $R_G = 1.30 - 1.35$ still leaves a large parameter space for outwards migration there. 

Overall we believe our results show that there is a significant possibility for the irradiation of the gap edge to severely slow the rate of Type II planetary migration. The results from our simplified approach can act as a good starting point for future three dimensional radiation-hydrodynamical simulations, which will reveal how important this effect is for allowing gas giant planets to remain at larger orbital radii.

\section*{Acknowledgements}
Sijme-Jan Paardekooper is supported by a Royal Society University Research Fellowship. We thank the reviewer David Tsang for his insightful comments.



\bibliographystyle{mnras}
\bibliography{mnrasRef}

\bsp	
\label{lastpage}
\end{document}